\documentclass[useAMS,usenatbib]{mn2e}

\usepackage{graphicx}
\usepackage{times}
\usepackage{natbib}
\usepackage{amsmath}

\newcommand{\gtrsim}{\ga}
\newcommand{\lesssim}{\la}

\def\zsun{{\rm Z_\odot}}
\def\msun{{\rm M_\odot}}
\def\msunh{{\rm M_\odot/{\it h}}}
\def\fnl{$f_{\rm NL}$}
\def\gnl{$g_{\rm NL}$}

\def\Mpch{{Mpc/{\it h}}}
\def\kpch{{kpc/{\it h}}}
\title[Baryon history and non-Gaussianities]{Baryon history and cosmic star formation in non-Gaussian cosmological models: numerical simulations}
\author[U.~Maio and F.~Iannuzzi]{
Umberto~Maio$^{1}$\thanks{E-mail: umaio@mpe.mpg.de},
Francesca~Iannuzzi$^{2}$\\
${}^1$Max-Planck-Institut f\"ur extraterrestrische Physik,
Giessenbachstra{\ss}e 1,  D-85748 Garching b. M\"unchen, Germany\\
${}^2$Max-Planck-Institut f\"ur Astrophysik,
Karl-Schwarzschild-Stra{\ss}e 1, D-85748 Garching b. M\"unchen, Germany\\
}

\begin{document}

\date{(draft)}
\pagerange{\pageref{firstpage}--\pageref{lastpage}}\pubyear{0}
\maketitle
\label{firstpage}

\begin{abstract}
We present the first numerical, N-body, hydrodynamical, chemical simulations of cosmic structure formation in the framework of non-Gaussian models.
We study the impact of primordial non-Gaussianities on early chemistry (e$^-$, H, H$^+$, H$^-$, He, He$^+$, He$^{++}$, H$_2$, H$_2^+$, D, D$^+$, HD, HeH$^+$), molecular and atomic gas cooling, star formation, metal (C, O, Si, Fe, Mg, S) enrichment, population III (popIII) and population II-I (popII) transition, and on the evolution of ``visible'' objects.
\\
We find that non-Gaussianities can have some consequences on baryonic structure formation at very early epochs, but the subsequent evolution at later times washes out any difference among the various models.
When assuming reasonable values for primordial non-Gaussian perturbations, it turns out that they are responsible for:
(i) altering early molecular fractions in the cold, dense gas phase of $\sim 10$ per cent;
(ii) inducing small temperature fluctuations of $\lesssim 10$ per cent during the cosmic evolution of primordial objects;
(iii) influencing the onset of the first star formation events, at $z>15$, and of the popIII/popII transition of up to some $10^7\,\rm yr$;
(iv) determining variations of $\lesssim 10$ per cent in the gas cloud and stellar mass distributions after the formation of the first structures;
(v) causing only mild variations in the chemical history of the Universe.
We stress, though, that purely non-Gaussian effects might be difficult to address, since they are strictly twisted with additional physical phenomena (e.g. primordial gas bulk flows, unknown primordial popIII stellar mass function, etc.) that have similar or stronger impact on the behaviour of the baryons.
\end{abstract}

\begin{keywords}
cosmology: theory -- structure formation
\end{keywords}


\section{Introduction}\label{Sect:introduction}
Cosmic structures are supposed to rise from primordial matter fluctuations generated at very early times, during the inflation era.
These fluctuations represent the seeds that grew over the cosmic time till the formation of the presently observed Universe \cite[e.g.][]{Komatsu2011}.
In the standard scenario, a Gaussian distribution of such perturbations is assumed, as a consequence of the central limit theorem.
\\
However, cosmic microwave background (CMB) temperature fluctuations, at high-order perturbation theory \cite[see e.g.][]{Zaldarriaga2000,Bartolo2004,Seery2005,Assadullahi2007,Cooray2008,Beltran2008,Liguori2008,Khatri2009,Lam2009,Bartolo2010,Yadav2010,Gao2010,Pitrou2010}, could have deviations from the purely Gaussian shape and leave room for non-Gaussian models.
Additional studies of the excursion set formalism and analyses of N-body numerical simulations of large-scale structures have shown how the abundance of rare-peak dark-matter haloes and their clustering properties and distribution might be affected \cite[e.g.][]{Grinstein1986,Koyama1999,Robinson2000,Grossi2007,Kang2007,Dalal2008,Grossi2009,Desjacques2009,Lam2010,Maggiore2010,Pillepich2010,Roncarelli2010,Pace2010,Wagner2010,LoVerde2011arXiv,Maturi2011arXiv,Yokoyama2011arXiv}.
\\
Observationally, the CMB is the best way to probe non-Gaussianities and different experiments (COBE, BOOMERanG, WMAP, PLANCK) can be used \cite[e.g.][]{Komatsu2002,Komatsu2003,Gaztanaga2003,Spergel2007,Hikage2008,Yadav2008,Afshordi2008,Slosar2008,Komatsu2009,Natoli2009,Komatsu2010,Hou2010,Raeth2009,Raeth2010arXiv}.
Recent determinations of the cosmological parameters by the 7-year Wilkinson Microwave Anysotropy Probe (WMAP) satellite \cite[e.g.][]{Komatsu2011} still find primordial non-Gaussianities.\\
Despite the huge amount of work based on N-body dark-matter only simulations, detailed chemical and hydrodynamical investigations of the effects on the visible matter are still missing, or poorly understood.
According to previous numerical studies \cite[][]{Viel2009}, the Lyman-$\alpha$ flux bispectrum in the high-redshift intergalactic medium is expected to present deviations from the Gaussian case up to $\sim 10$ per cent at $z\sim 4$.
Semianalytic estimates \cite[e.g.][]{Crociani2009} seem to show that the ionized fraction of the intergalactic medium (IGM) can grow by a factor of a few, up to 5, with respect to the corresponding Gaussian model.
The increase of the filling factor has a small impact on the reionization optical depth and is of the order of $\sim 10$ per cent if a scale-dependent non-Gaussianity is assumed.
Since non-Gaussianities are expected to influence the primordial matter density fluctuations, it is important to address their consequences on the history of the baryons, from the early ``dark ages'', when stars were not formed, yet, to the following ``louminous ages'', after the birth of the first objects.
Additionally, it is relevant to understand the role they play in the transition from the primordial population III (popIII) stars to present-day population II-I (popII) stars.
These are indeed among the goals of the present work.
\\
A standard way to account for the existence of non-Gaussianities in the primordial fluctuation field is to regard them as a perturbation on a Gaussian background. Strongly non-Gaussian models have been progressively dismissed by CMB observations and it is now generally agreed that primordial non-Gaussianities, if present, must have represented a moderate deviation from the underlying Gaussian statistics. The most commonly adopted parametrisation of non-Gaussianities considers them as second-order perturbations of the Bardeen gauge-invariant potential \cite[see e.g.][]{Salopek1990,Komatsu2001,Verde2010,Desjacques2010}:
\begin{equation}\label{eq:nong}
\Phi = \Phi_{\rm L} + f_{\rm NL} \left[ \Phi_{\rm L}^2 - <\Phi_{\rm L}^2> \right],
\end{equation}
where $\Phi_{\rm L}$ is the {\it linear} Gaussian part, and the dimensionless coupling constant, \fnl, controls the importance of the non-Gaussian contribution\footnote{
The Bardeen gauge-invariant potential, $\Phi$, reduces to the usual Newtonian peculiar gravitational potential, up to a minus sign, on sub-Hubble scales.
In the literature \cite[e.g.][]{Verde2010}, there are two, sometimes misleading, notations: the large-scale structure (LSS) convention and the CMB convention.
In the former (LSS), $\Phi$ is linearly extrapolated at $z=0$;
in the latter (CMB), $\Phi$ denotes the primordial potential, at $z\rightarrow +\infty$.
The net difference is related to the ratio of the growth factor, $g(z)$, at the two times and accounts for a correction of roughly unity:
i.e.
\fnl$^{\rm LSS}$ 
$= g(z\rightarrow +\infty)/g(z=0)\, $
\fnl$^{\rm CMB}$,
which corresponds to $ \sim 1.28$\fnl$^{\rm CMB}$, for a standard $\Lambda$CDM model,
and to $ \sim 1.33$\fnl$^{\rm CMB}$, for the latest observational data from WMAP \cite[][]{Komatsu2011}.
Usually, observational values are reported according to the CMB convention.
}.
Since the final value of the potential $\Phi$ depends on the local value of the Gaussian field $\Phi_{\rm L}$, non-Gaussianities of the the form (\ref{eq:nong}) are labelled as ``local''. Local models are physically relevant for scenarios where non-linearities develop outside the horizon, but they are by no means exhaustive of all the forms non-Gaussianities can assume according to the several mechanisms that generate them \cite[for further details see e.g][and references therein]{Babich2004,Bartolo2004,Loverde2008}.
The exact value of the \fnl{} parameter is continuously better constrained by CMB observations (see discussion in Sect.~\ref{Sect:discussion}).
More extended analyses could in principle be done by introducing additional parameters, like the cubic-order one, \gnl, but, given their large uncertainties \cite[e.g. $-7.4\times 10^5<$\gnl$<8.2\times 10^5$, according to][]{Smidt2010}, we will consider only second-order corrections.
In addition, since latest determinations suggest \fnl$\gtrsim 0$ \cite[][]{Spergel2007,Slosar2008,Afshordi2008,Yadav2008,Curto2009b,Komatsu2009,Smith2009,Rudjord2009,Komatsu2010,Hou2010,Rudjord2010,Cabella2010}, we will only consider positive values for \fnl.
\\
In the present work we will focus on the effects non-Gaussianities have on the baryon history in universes with different initial conditions, by studying how gas evolution, cooling and condensation could lead to different star formation epochs for different initial scenarios, and by exploring how the various stellar population regimes and the statistical properties of gas and stars are affected in non-Gaussian models.\\
We will present the simulations performed in Sect.~\ref{Sect:simulations}, 
together with the main results, in Sect.~\ref{Sect:results}, about 
the first cosmic structures (Sect.~\ref{Sect:maps}), star formation (Sect.~\ref{Sect:SFR}), chemistry evolution (Sect.~\ref{Sect:chemistry}) and primordial streaming motions \cite[see also e.g.][]{TseliakhovichHirata2010,Maio2011}, and baryon statistics (Sect.~\ref{Sect:distributions}).
Then, we will discuss and conclude in Sect.~\ref{Sect:discussion}.

\section{Simulations}\label{Sect:simulations}
\begin{table*}
\centering
\caption[Simulation set-up]{Initial parameters for the different runs.}
\begin{tabular}{lccccccc}
\hline
\hline
Runs& Box side & Particle mass [$\msun/h$] for & Softening    &\fnl & Gas bulk        & PopIII IMF\\
    & [\Mpch]  & gas (dark matter)             & [kpc/{\it h}]&     & velocity [km/s] & range [M$_\odot$]\\
\hline
Run05.0    & 0.5  & $42.35\quad$ ($275.28$)& 0.04 & 0    & 0 & [100, 500]\\
Run05.10   & 0.5  & $42.35\quad$ ($275.28$)& 0.04 & 10   & 0 & [100, 500]\\
Run05.50   & 0.5  & $42.35\quad$ ($275.28$)& 0.04 & 50   & 0 & [100, 500]\\
Run05.100  & 0.5  & $42.35\quad$ ($275.28$)& 0.04 & 100  & 0 & [100, 500]\\
Run05.1000 & 0.5  & $42.35\quad$ ($275.28$)& 0.04 & 1000 & 0 & [100, 500]\\
\hline
Run100.0    & 100  & $3.39\times 10^8\quad$ ($2.20\times 10^9$)& 7.8 & 0    & 0 & [100, 500]\\
Run100.10   & 100  & $3.39\times 10^8\quad$ ($2.20\times 10^9$)& 7.8 & 10   & 0 & [100, 500]\\
Run100.50   & 100  & $3.39\times 10^8\quad$ ($2.20\times 10^9$)& 7.8 & 50   & 0 & [100, 500]\\
Run100.100  & 100  & $3.39\times 10^8\quad$ ($2.20\times 10^9$)& 7.8 & 100  & 0 & [100, 500]\\
Run100.1000 & 100  & $3.39\times 10^8\quad$ ($2.20\times 10^9$)& 7.8 & 1000 & 0 & [100, 500]\\
\hline
Run100.0.SL    & 100  & $3.39\times 10^8\quad$ ($2.20\times 10^9$)& 7.8 & 0    & 0 & [0.1, 100]\\
Run100.10.SL   & 100  & $3.39\times 10^8\quad$ ($2.20\times 10^9$)& 7.8 & 10   & 0 & [0.1, 100]\\
Run100.50.SL   & 100  & $3.39\times 10^8\quad$ ($2.20\times 10^9$)& 7.8 & 50   & 0 & [0.1, 100]\\
Run100.100.SL  & 100  & $3.39\times 10^8\quad$ ($2.20\times 10^9$)& 7.8 & 100  & 0 & [0.1, 100]\\
Run100.1000.SL & 100  & $3.39\times 10^8\quad$ ($2.20\times 10^9$)& 7.8 & 1000 & 0 & [0.1, 100]\\
\hline
Run05.0.30 & 0.5  & $42.35\quad$ ($275.28$)& 0.04 & 0    & 30 & [100, 500]\\
Run05.0.60 & 0.5  & $42.35\quad$ ($275.28$)& 0.04 & 0    & 60 & [100, 500]\\
Run05.0.90 & 0.5  & $42.35\quad$ ($275.28$)& 0.04 & 0    & 90 & [100, 500]\\
Run05.100.30 & 0.5  & $42.35\quad$ ($275.28$)& 0.04 & 100    & 30 & [100, 500]\\
Run05.100.60 & 0.5  & $42.35\quad$ ($275.28$)& 0.04 & 100    & 60 & [100, 500]\\
Run05.100.90 & 0.5  & $42.35\quad$ ($275.28$)& 0.04 & 100    & 90 & [100, 500]\\
\hline
\label{tab:runs}
\end{tabular}
\begin{flushleft}
\vspace{-0.5cm}
{\small
}
\end{flushleft}
\end{table*}
We use a modified version of the parallel tree/SPH numerical code P-Gadget2 \cite[][]{Springel2005}.
Beyond gravity and hydrodynamics, it includes radiative gas cooling at high \cite[][]{SD1993} and low \cite[][]{Maio2007} temperatures, both from molecules and fine-structure metal transitions, multiphase model \cite[][]{Springel2003} for star formation \cite[inspired on the works by][]{KatzGunn1991,Cen1992,CenOstriker1992,Katz1992,Katz_et_al_1992,Katz_et_al_1996}, UV background radiation \cite[][]{HaardtMadau1996}, and wind feedback \cite[][]{Springel2003,Aguirre_et_al_2001}.
\\
The main agents of gas cooling, in hot environments, are atomic resonant transitions of H and He excitations that can rapidly bring the temperatures down to $\sim 10^4\,\rm K$.
In metal-rich environments, atomic fine-structure transitions can further lower them \cite[see][]{Maio2007}, mostly at high redshift, since the formation of a UV background \cite[][]{HaardtMadau1996,Haehnelt2001,Gilmore2009} at lower redshifts competes by heating the IGM.
In dense environments, stochastic star formation is assumed, cold gas is gradually converted into stars (that reproduce the Kennicut-Schmidt low), outflows take place, and feedback effects inject entropy into the surrounding environment, leading to a self-regulated star forming regime \cite[for further detail see ][]{Springel2003}.
\\
In the code, we also include the relevant chemical network to self-consistently follow the cosmic evolution of e$^-$, H, H$^+$, H$^-$, He, He$^+$, He$^{++}$, H$_2$, H$_2^+$, D, D$^+$, HD, HeH$^+$ \cite[see e.g.][ and references therein]{Yoshida2003,Maio2007,Maio2009,Maio2010a}, metal (C, O, Si, Fe, Mg, S) pollution from popIII and/or popII stellar generations, ruled by a critical metallicity threshold of $Z_{crit}=10^{-4}\,\zsun$ \cite[][]{Tornatore2007,Maio2010a} and leading gravitational enrichment into the surrounding medium \cite[][]{Maio2011arXiv}.
\\
We perform several runs (see Tab.~\ref{tab:runs}) in order to check the different assumptions on \fnl, at large and small scales.
\\
The initial conditions are generated with a modified version of the N-GenIC code that introduces non-Gaussian features of the local type in the initial density field according to eq. (\ref{eq:nong}) \cite[for further details see e.g.][]{Bartolo2005,Grossi2007,Grossi2009,Viel2009}.
We assume a concordance $\Lambda$CDM model with
matter density parameter $\Omega_{\rm 0,m}=0.3$,
cosmological density parameter  $\Omega_{\rm 0,\Lambda}=0.7$,
baryon density parameter $\Omega_{\rm 0,b}=0.04$,
expansion rate at the present of H$_0=70\,\rm km/s/Mpc$,
power spectrum normalization via mass variance within 8~Mpc/{\it h} radius sphere $\sigma_8=0.9$,
and spectral index $n=1$.
We study the non-Gaussianity parameter space identified by \fnl = 0, 10, 50, 100, 1000, in cosmological periodic boxes of 0.5~\Mpch{ } and 100~\Mpch{} a side, sampled, at redshift $z=100$, with $2\times 320$ particles with gas mass resolution of $\sim 40\,\msunh$ and $\sim 3\times 10^8\,\msunh$, and corresponding maximum physical softening lengths of 0.04~\kpch{} and 7.8~\kpch, respectively.
This will allow us to see the impacts of the different models down to scales of the order of $\sim 0.1\,\rm kpc$.
We note that the small size\footnote{
Results on biased high-density (zoomed) initial conditions, for which the onset of star formation is expected to take place at redshifts as high as $\sim 40-50$ heve been presented by e.g. \cite{Maio2009}.
While in such set-ups the timing of the popIII formation might be more accurate and independent from Lyman-Werner radiation feedback \cite[][]{Johnson2008,TrentiStiavelli2009}, the general physical trends are supposed to be very similar.
}
of the 0.5~\Mpch{} boxes prevents the simulations from being run to very low redshifts, in order for the fundamental mode to remain linear; the large simulations satisfy instead this requirement \cite[see e.g. Sect.~2.3 in][]{Maio2009PhDT} and are thus run until $z=0$.
The simulations will be labelled as Run05.0, Run05.10, Run05.50,Run05.100, Run05.1000, and Run100.0, Run100.10, Run100.50,Run100.100, Run100.1000, respectively.\\
Throughout this work, we will assume a Salpeter IMF for popII star forming regions (where $Z\ge Z_{crit}$), and a top-heavy IMF, with mass range [100$\msun$, 500$\msun$], for popIII star forming regions (where $Z<Z_{crit}$).
The latter assumption is matter of debate, since there are also theoretical and numerical evidences for the existence of low-mass popIII stars \cite[e.g.][]{Yoshida2006, Yoshida_et_al_2007,CampbellLattanzio2008,SudaFujimoto2010,Stacy2010}.
If this is the case, that might mean that the conclusions of several previous works predicting massive primordial stars were highly affected by numerics and resolution.
As the primordial popIII IMF can impact the cosmological pollution history \cite[e.g.][]{Maio2011arXiv}, in Sect.~\ref{Sect:SFR}, we we will also consider, for the 100~\Mpch{ } side boxes, the opposite extreme of a Salpeter-like popIII distribution, with appropriate yields, for primordial stars \cite[see discussion in][and references therein]{Maio2010a}, and mass range [0.1$\msun$, 100$\msun$]: the corresponding runs will be labelled by adding the SL suffix.\\
In order to see how relevant the consequences of non-Gaussian perturbations are,  in Sect.~\ref{Sect:SFR}, we will also compare them with second-order effects in the linearized perturbation theory \cite[][]{TseliakhovichHirata2010}, that expect supersonic gas bulk flows at early times to suppress structure growth, and delay star formation on scales smaller than a few Mpc \cite[][]{Maio2011}.
We will focus on the \fnl=0 and \fnl=100 cases, assuming primordial bulk shifts of 30, 60, 90~km/s.
The six corresponding simulations will be identified by Run05.0.30, Run05.0.60, Run05.0.90 (for the \fnl=0 set), and by Run05.100.30, Run05.100.60, Run05.100.90 (for the \fnl=100 set), respectively.\\
A friend-of-friend (FoF) algorithm \cite[][]{Springel2001}, with comoving linking lenght of 20 per cent the mean inter-particle separation, is applied at postprocessing time, considering all the types of particles, to find the formed cosmic objects, with their dark, gaseous, and stellar components.\\
We summarize the properties of all the different runs in Table~\ref{tab:runs}.


\section{Results}\label{Sect:results}

In the following, we will show the main repercussions primordial non-Gaussianities have on the cosmic baryon history in the Universe.
We will focus on cosmic structure evolution (Sect.~\ref{Sect:maps}), star formation (Sect.~\ref{Sect:SFR}), chemical properties of the Universe (Sect.~\ref{Sect:chemistry}), and baryon statistics (Sect.~\ref{Sect:distributions}) in different models.

\begin{figure*}
\centering
\includegraphics[width=\textwidth]{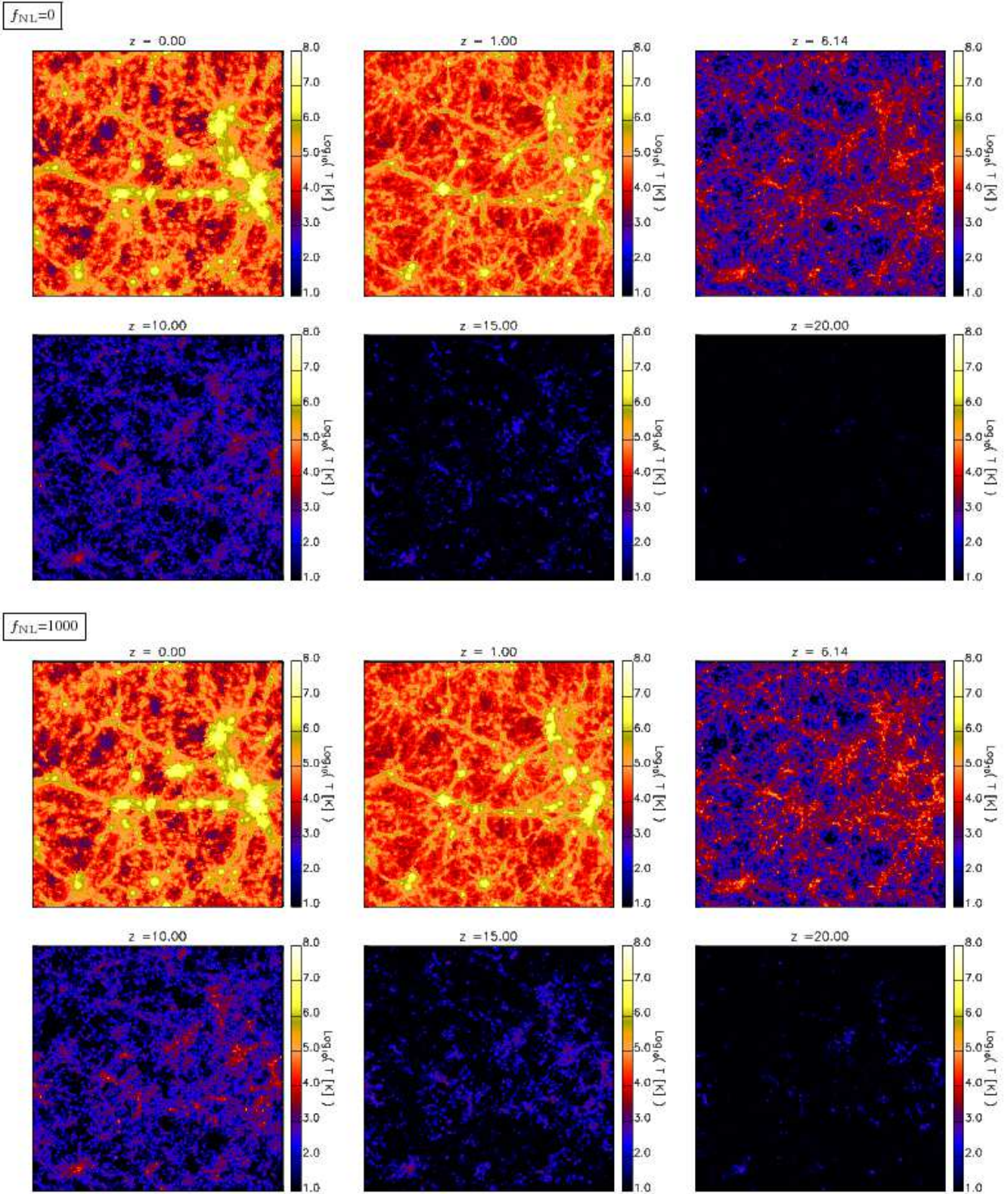}
\caption[Maps]{\small
Mass-weighted temperature evolution for the large-scale 100~\Mpch{} side boxes, with \fnl=0 (first and second row), and \fnl=1000 (third and fourh row), at different redshifts (see legends).
All the maps refer to slices centered at the mid-plane of the box, height $z=50\,\rm\Mpch$, and have thickness of $\sim 7\,\rm\Mpch$, i.e. $1/14$ the boxsize.
}
\label{fig:Maps}
\end{figure*}

\subsection{Structure evolution}\label{Sect:maps}
\begin{figure*}
\centering
\includegraphics[width=\textwidth]{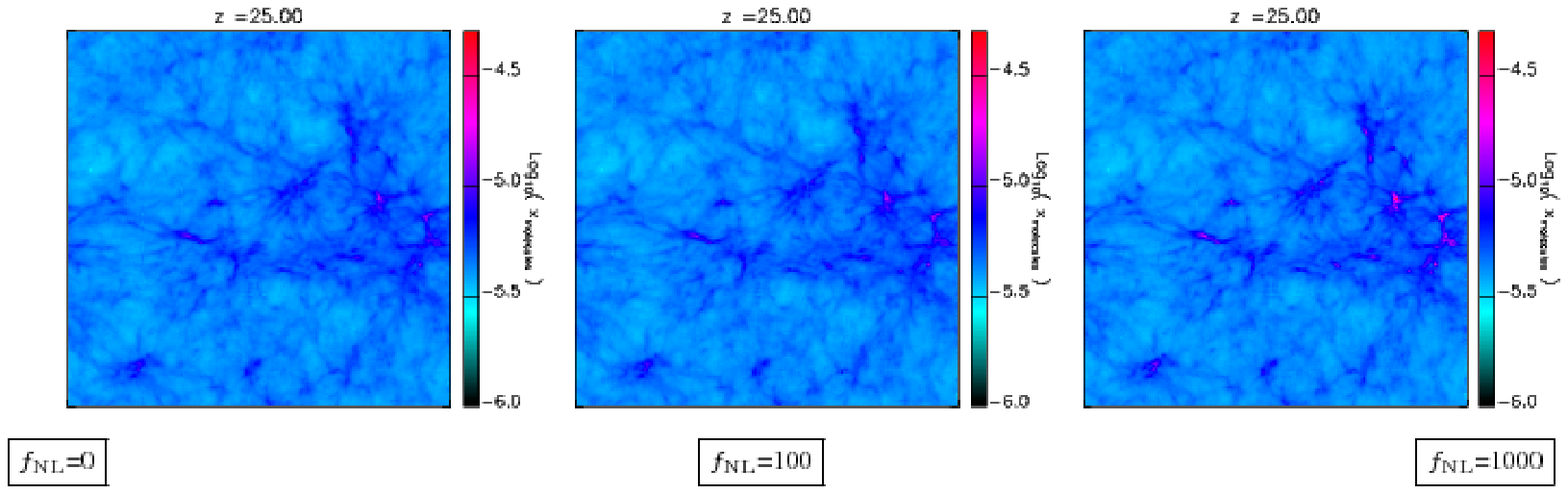}
\\
\caption[Maps2]{\small
Molecule maps at redshift $z=25$ for the high-resolution 0.5~\Mpch side boxes, with \fnl=0 (left), \fnl=100 (center), and \fnl=1000 (right).
All the maps refer to slices centered at the mid-plane of the box, height $z=0.25\,\rm\Mpch$, and have thickness of $\sim 0.03\,\rm\Mpch$, i.e. $1/14$ the boxsize.
}
\label{fig:Maps2}
\end{figure*}

\begin{figure}
\centering
\includegraphics[width=0.33\textwidth]{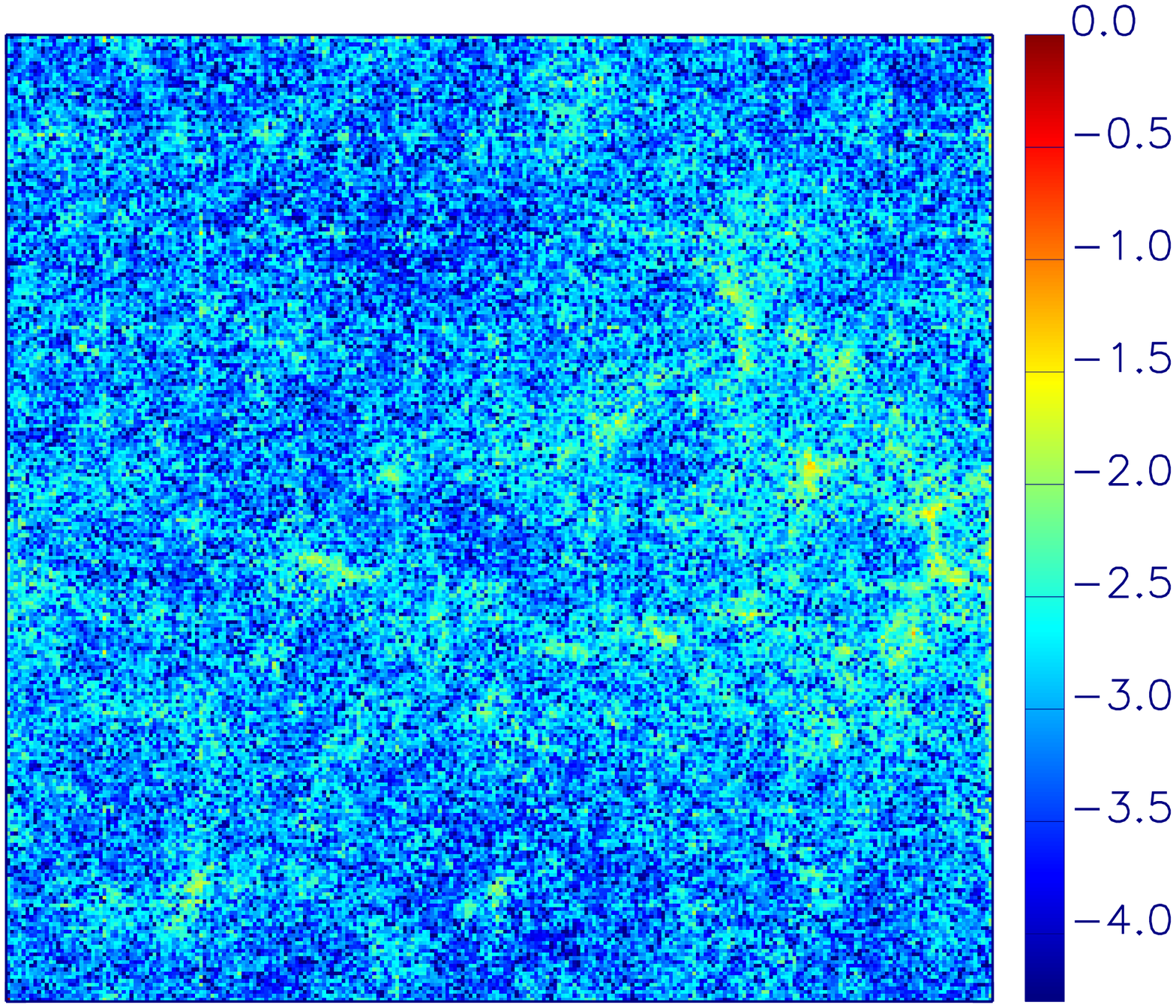}\\
\fbox{\fnl=100 -- \fnl=0}\\
\includegraphics[width=0.33\textwidth]{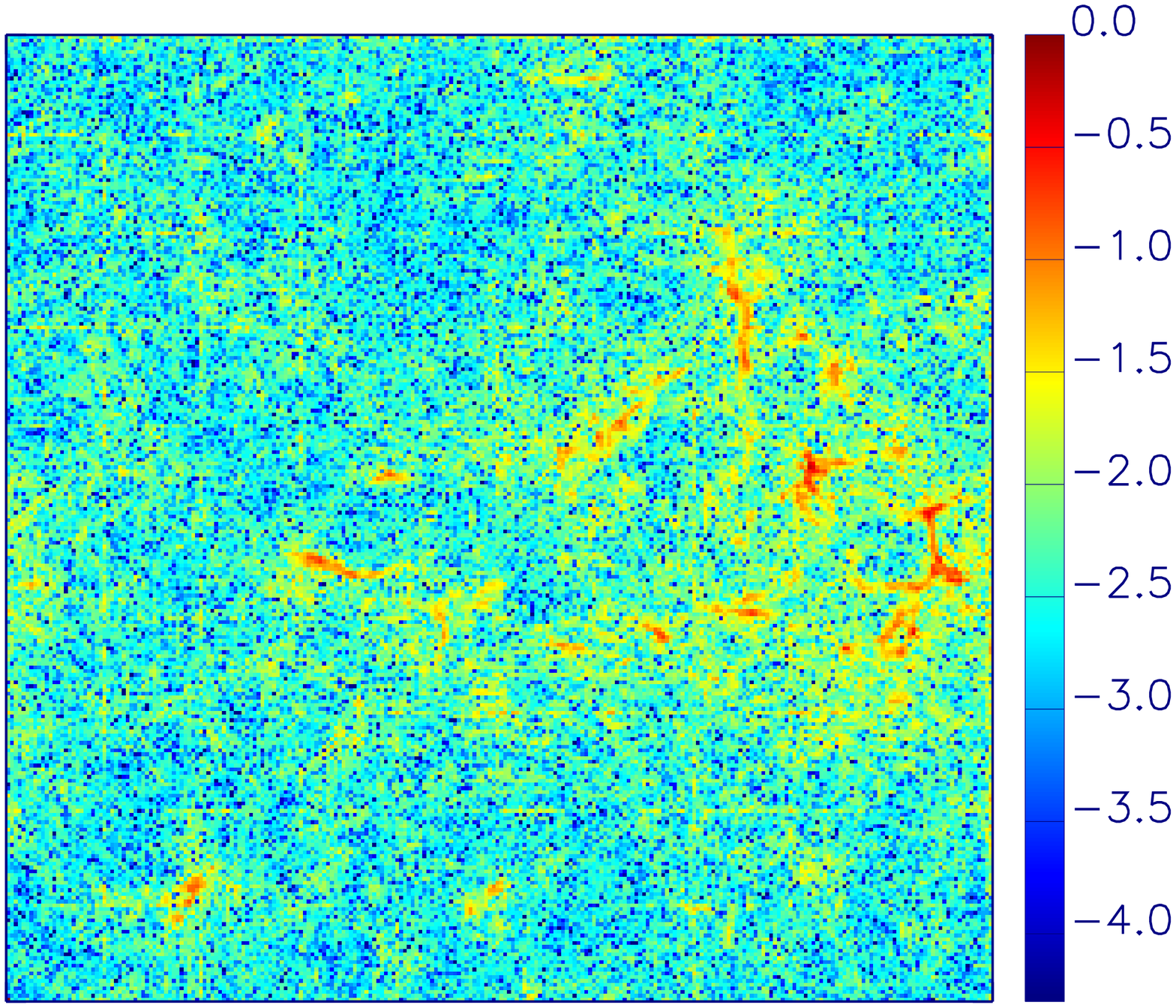}\\
\fbox{\fnl=1000 -- \fnl=0}\\
\caption[Maps3]{\small
Molecular fraction contrast at redshift $z=25$ between the \fnl=100 and \fnl=0 cases (top), and between the \fnl=1000 and \fnl=0 cases (bottom).
The color scales refer to the logarithm of the absolute value of the molecular fraction contrasts in the two cases.
}
\label{fig:Maps3}
\end{figure}

We start by showing the basic features of the simulations.
In Fig.~\ref{fig:Maps}, we give a pictorial representation of the simulation with side of 100~\Mpch, and display the mass-weighted temperature evolution for a slice centered in the middle of the box, along the z-direction, having a thickness of $\sim 7\,\rm\Mpch$, and smoothed over a grid of 1024$\times$1024 pixels.
We consider the \fnl=0 (first two rows) and the \fnl=1000 (last two rows) cases, and highlight the differences among them.
At high redshift, $z\gtrsim 20$, the \fnl=1000 case shows more perturbations with respect to the \fnl=0 one, and this is visible at $z\sim 10-15$, as well, when feedback effects from early star formation heat the medium above $\sim 10^4-10^5\,\rm K$.
The maps of the \fnl=1000 case present more advanced stages of structure formation, and this keeps the gas slightly hotter, with temperatures of $\sim 2\times 10^6\,\rm K$, until $z\sim 6$, i.e. for about the first Gyr of the Universe.
Temperature variations are up to a factor of $\lesssim 2$ for the \fnl=1000 case, but only up to $\sim 10$ per cent, for the other ones.
Later on, at $z<6$, more differences are not clearly visible and the final trends at lower redshift ($z\sim 1$) catch up and lead to the same behaviour at $z\sim 0$.\\
We conclude that baryon history is affected by primordial non-Gaussianities mostly at early times, and this statement is strengthened from Fig.~\ref{fig:Maps2}, where we plot the molecular fraction at redshift $z=25$, for the high-resolution runs, with side of 0.5~\Mpch.
We compare the \fnl=0, the \fnl=100, and the \fnl=1000 cases during the very first stages of structure formation via molecular (mainly H$_2$ and HD) cooling and catastrophyc run-away collapse.
While in low-density environments there are no strong net differences, in the high-density regions the molecular content can change up to a factor of a few.
Indeed, in the \fnl=100 case, molecules increase of $\sim 10$ per cent (more exactly of 6 per cent) with respect to the \fnl=0 one, and in the \fnl=1000 case, of a factor of $\sim 2.5$.
\\
In Fig.~\ref{fig:Maps3}, we highlight the differences by plotting the contrast between the \fnl=100 and \fnl= 0 cases (top), and between the \fnl=1000 and \fnl=0 cases (bottom), defined, for each pixel in the map, as
\begin{equation}
C_{\rm molecules} = \frac{x_{\rm molecules, 1}-x_{\rm molecules, 0}}{x_{\rm molecules, 1}+x_{\rm molecules, 0}}.
\end{equation}
In the previous equation, $x_{\rm molecules, 1}$ is the molecular fraction in the pixel for the \fnl=100 or the \fnl=1000 case; while $x_{\rm molecules, 0}$ is the reference \fnl=0 case.
It is clear that the comparison between \fnl=100 and \fnl=0 expects much smaller differences than the one between \fnl=1000 and \fnl=0.
Indeed, the top panel presents about 54 per cent of the entire area corresponding to contrasts of $\lesssim 10^{-3}$ (blue), almost 45 per cent corresponding to contrasts of $\sim 10^{-3}-10^{-2}$ (cyan-green), and the remaining few per cent reaching values of $\sim 10^{-1}$, or larger (yellow-red).
On the other side, the bottom panel displays larger deviations from the reference \fnl=0 case, with only 15 per cent of the region having contrasts $\lesssim 10^{-3}$ (blue), roughly 70 per cent of $\sim 10^{-3}-10^{-2}$ (cyan-green), and the remaining 15 per cent larger values (yellow-red).
This means that larger \fnl{} will induce higher molecular production, stronger cooling, earlier gas collapse and earlier star formation, as we will see in the next section.
\\
The fundamental reason why gas and baryonic structures are sensitive to different \fnl{} is that primordial non-Gaussianities affect the growth and evolution of the underlying dark matter haloes.
As a consequence, the resulting baryon history reflects the imprints of the primordial dark-matter non-Gaussian distribution.


\subsection{Star formation}\label{Sect:SFR}
\begin{figure*}
\centering
\includegraphics[width=0.45\textwidth]{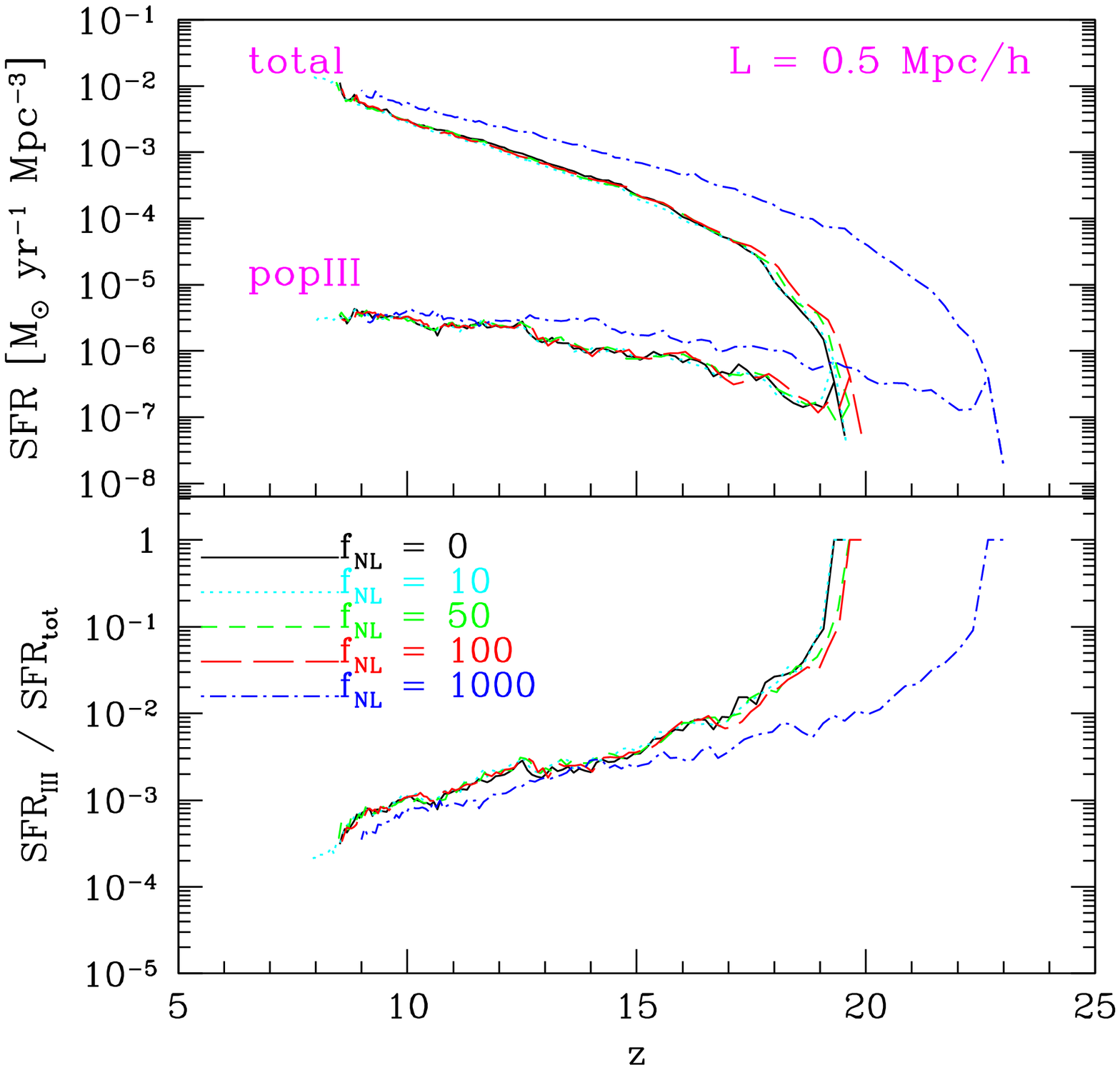}
\includegraphics[width=0.45\textwidth]{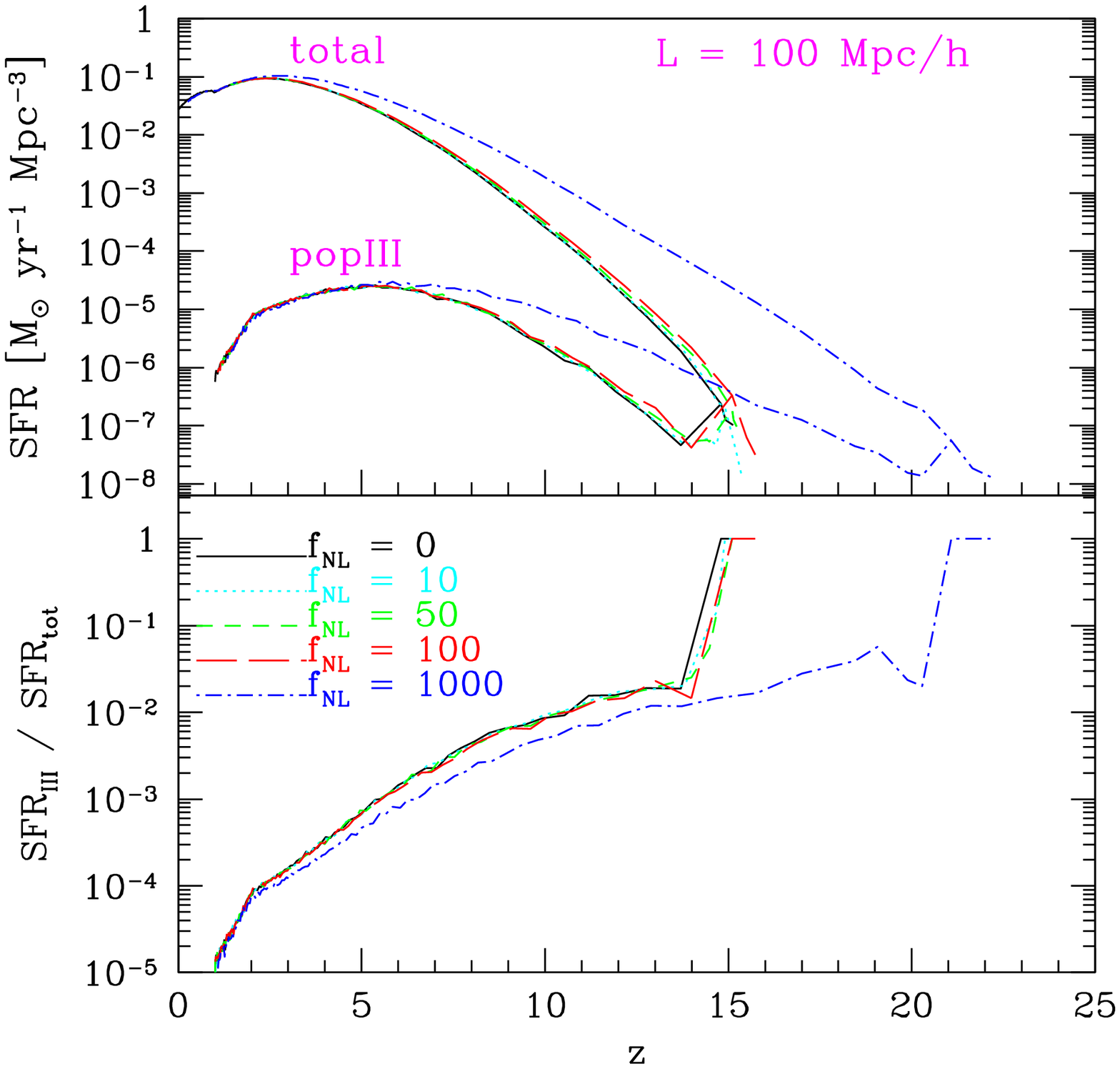}\\
\caption[SFR]{\small
Star formation rate densities as a function of the redshift for the simulation runs of the 0.5~\Mpch{} side boxes (left) and 100~\Mpch{} side boxes (right).
In the top panels, the upper lines refer to the total star formation rate densities, while the bottom lines to the popIII ones.
In all the cases, we consider
\fnl=0 (black solid lines),
\fnl=10 (cyan dotted lines),
\fnl=50 (green short-dashed lines),
\fnl=100 (red long-dashed lines), and
\fnl=1000 (blue dotted short-dashed lines.
They correspond to Run05.0, Run05.10, Run05.50, Run05.100, Run05.1000, on the left panels,
and to Run100.0, Run100.10, Run100.50, Run100.100, Run100.1000, on the right panels.
Below, we plot the corresponding popIII contributions to the total star formation rates.
}
\label{fig:SFR}
\end{figure*}
\begin{figure}
\centering
\includegraphics[width=0.45\textwidth]{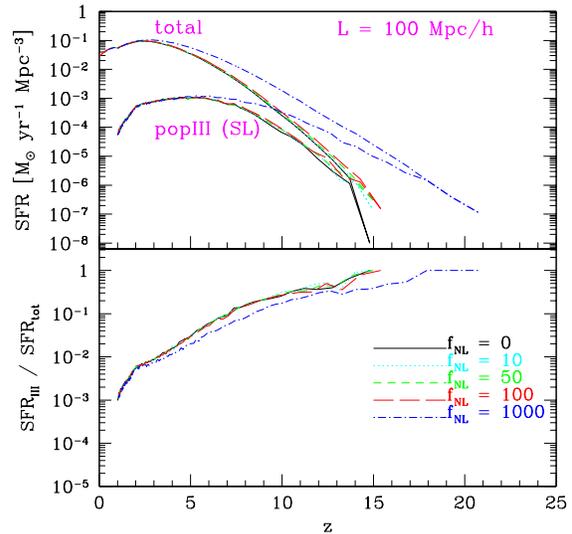}
\caption[SFR Salpeter PopIII IMF]{\small
Star formation rate densities as a function of the redshift for the simulations of the 100~\Mpch{} side boxes, with Salpeter popIII IMF.
In the top panels, the upper lines refer to the total star formation rate densities, while the bottom lines to the popIII ones.
In all the cases, we consider
\fnl=0 (black solid lines),
\fnl=10 (cyan dotted lines),
\fnl=50 (green short-dashed lines),
\fnl=100 (red long-dashed lines), and
\fnl=1000 (blue dotted short-dashed lines.
They correspond to Run100.0.SL, Run100.10.SL, Run100.50.SL, Run100.100.SL, Run100.1000.SL.
Below, we plot the corresponding popIII contributions to the total star formation rates.
}
\label{fig:SFR_IMFrange}
\end{figure}
\begin{figure}
\centering
\includegraphics[width=0.45\textwidth]{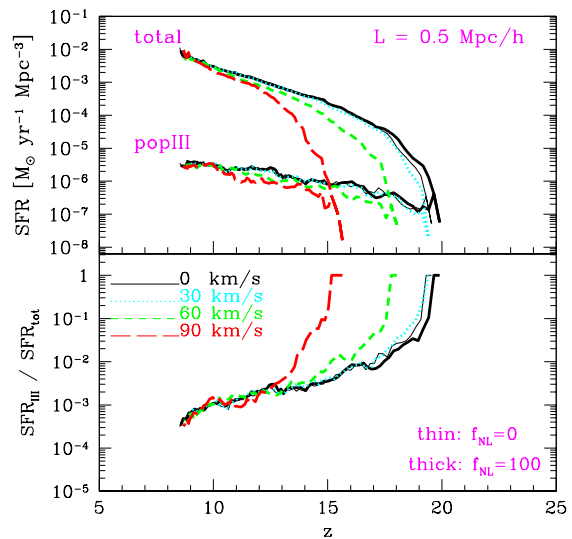}
\caption[vb]{\small
Upper panel: star formation rate densities as a function of the redshift for the simulation runs of the 0.5~\Mpch{} side boxes with \fnl=100, and primordial gas bulk velocities of
0~km/s (thick black solid line),
30~km/s (thick cyan dotted lines),
60~km/s (thick green short-dashed lines), and
90~km/s (thick red long-dashed lines).
As a comparison, we overplot the results for the 0.5~\Mpch{} side box with \fnl=0 and 0~km/s for the primordial gas bulk velocity (thin black solid line).
Bottom panel: the corresponding popIII contributions to the total star formation rates.
}
\label{fig:SFR_vb}
\end{figure}

At this point, we study the star formation history in the different models, paying attention to both the popIII and the popII regime.\\
In Fig.~\ref{fig:SFR}, we plot the star formation rate densities from the simulations of the 0.5~\Mpch{} side boxes (left), and 100~\Mpch{} side boxes (right), with the relative popIII contribution (lower panels).
The simulations have similar trends, but the onset of star formation is slightly earlier for the runs in 0.5~\Mpch{} side boxes.
This is essentially due to the fact that the higher-resolution simulations are able to capture better molecular cooling and gas collapse in primordial mini-haloes.
Thus, they can address star formation in those regimes (i.e. at masses smaller than $\sim 10^{8}\msunh$) where the larger 100~\Mpch{} side boxes are limited by resolution (see properties in Tab.~\ref{tab:runs}).
We note that at both small and large scales, the main effect of non-Gaussianities is a general shift of star formation, for the two population regimes.
This is due to the fact that at larger \fnl{} the contribution to the exponential tail of the mass function increases, leading to higher and earlier star formation.
The differences among the various onsets are very small for \fnl=0, \fnl=10, \fnl=50 and \fnl=100, and correspond to some $\sim 10^7\,\rm yr$.
Only for the \fnl=1000 case there is a larger gap of $\Delta z\sim$ a few, i.e. $\sim 5\times 10^7\,\rm yr$, in the 0.5~\Mpch{} box, and $\sim 10^8\,\rm yr$, in the 100~\Mpch{} box.
The behaviours of the popIII regimes are very little affected and their overall contributions drop down to $\sim 10^{-3} - 10^{-2}$, for any \fnl, by redshift $z\sim 10$.
Due to the anticipated evolution of the runs with larger \fnl, the corresponding popIII contributions drop down earlier and stay slightly below the \fnl=0 cases.
On small scales (left panels), the difference is about $\sim 2$ orders of magnitude, at $z\sim 19-20$, but only a factor of a few at later stages.
At $z\sim 10$, in the \fnl=0 case one expects a popIII contribution of $\sim 0.0010$, while in the \fnl=1000 case, of $\sim 0.0008$ -- 20 per cent smaller.
Similarly, at the same redshift, on larger scales (righ panels), the differences are well within a factor of 2, with a popIII contribution of $\sim 0.0097$ for \fnl=0, and $\sim 0.0055$ for \fnl=1000.
\\
To check for degeneracies among different physical processes involved in the description of cosmic structure evolution, we have studied also how the scenario changes in two additional cases:
\begin{itemize}
\item
  by adopting a common Salpeter-like IMF for the primordial popIII generation;
\item
  by comparing with second-order corrections to the linear perturbation theory \cite[][]{TseliakhovichHirata2010}, which are supposed to generate primordial gas bulk motions, to suppress star formation on scales smaller than a few Mpc \cite[][]{Maio2011}, and to induce delays in the reionization of the Universe \cite[][]{Dalal2010}.
\end{itemize}
In Fig.~\ref{fig:SFR_IMFrange}, we plot the star formation rate densities for the 100~\Mpch{} side boxes with primordial \fnl=0, \fnl=10, \fnl=50, \fnl=100, and \fnl=1000.
In these cases (see details in Tab.~\ref{tab:runs}), we have considered a Salpeter-like popIII IMF that is expected to cause a longer popIII-dominated epoch, because of the loger lifetimes of the $\sim 10\rm s\,\msun$ stars dying as supernovae \cite[see][]{Maio2010a}.
In fact, we find that the popIII regime dominates at early times and its contribution drops down to $\sim 10$ per cent only after $\sim 4\times 10^8\,\rm yr$.
Furthermore, the global popIII contributions are higher in comparison with the top-heavy popIII IMF previously discussed, and the difference is more than one order of magnitude.
At redshift $z\sim 10$, it is roughly a few times $10^{-1}$, which means that the primordial IMF can have impacts on the baryon and stellar evolution that are much stronger than the non-Gaussian effects.
Therefore, if we want to give constraints on primordial non-Gaussianities it will be important first to properly constrain the popIII IMF.\\
About the second check, we use only the small-box simulations, since they are more suitable to take into account primordial streaming motions.
These are predicted to have rms velocities of the order of $\sim 30\,\rm km/s$ at decoupling  \cite[][]{TseliakhovichHirata2010}.
So, we focus (see details in Tab.~\ref{tab:runs}) on the same initial conditions described above for \fnl=0 and \fnl=100, and add \cite[as in][]{Maio2011} a primordial bulk shift of 30~km/s to the gas.
To consider statistical deviations from the rms value, we additionally run the cases with 60~km/s and 90~km/s shifts (i.e. 2 and 3 times the rms value), as upper limits, which can be found in rare regions (less than $\sim 1$ per cent) of the Universe, though.
In Fig.~\ref{fig:SFR_vb}, we compare the results for the two different \fnl, and plot the \fnl=0 and no-shift case (thin solid line) together with all the \fnl=100 cases (thick lines).
Obviously, star formation sets in at different times, according to the different values of the streaming motions and the resulting time delays are of the order of tens of Myr ($\Delta z\sim$ a few).
However, if we compare the \fnl=0 (thin solid line) and the \fnl=100 (thick lines) cases for a fixed bulk shift velocity, we notice that the effects of non-Gaussianities are comparable to or negligible with respect to the impacts of the streaming motions.
The star formation rate in the \fnl=100 and 30~km/s-shift (thick dotted line) case almost coincides with the one in the \fnl=0 run, and this leads to a degeneracy between these two different phenomena.
Moreover, in the \fnl=100 and 60~km/s-shift (thick short-dashed line) case and in the \fnl=100 and 90~km/s-shift (thick long-dashed line) case streaming motions cause much stronger delays.
The value of 30~km/s is the average expected in the whole Universe, but different values can be found in different regions, thus, purely non-Gaussian effects are going to be strongly twisted with the statistical variations of such second-order contaminations.
\\
Finally, we stress that variations on the determination of $\sigma_8$ of a few percents could produce a similar degeneracy, and different primordial Lyman-Werner (LW) radiation strengths coming from the early star forming episodes could probably have even larger effects.
In fact, the LW radiation (which is not explicitely followed in the simulations persented here) could dissociate primordial molecules, partially inhibiting H$_2$ cooling in the neighbouring sites.
Thus, the star formation process and the popIII formation at early times are expected to be delayed to lower redshifts.
But the metals which are then ejected from primordial stars are insensitive to LW radiation: independently from molecule formation, they strongly enhance the cooling capabilities of the surrounding gas at any temperature \cite[e.g.][]{Maio2007}, and can rapidly lead to a popII-dominated regime \cite[][]{Maio2010a,Maio2011}.
Given the tight connections between primordial IMF, LW radiation and chemical feedback more detailed studies will be needed to draw definitive conclusions on this issue, but definitely these will not change the results related to the \fnl{} effects.


\subsection{Chemical evolution}\label{Sect:chemistry}
\begin{figure}
\centering
\includegraphics[width=0.45\textwidth]{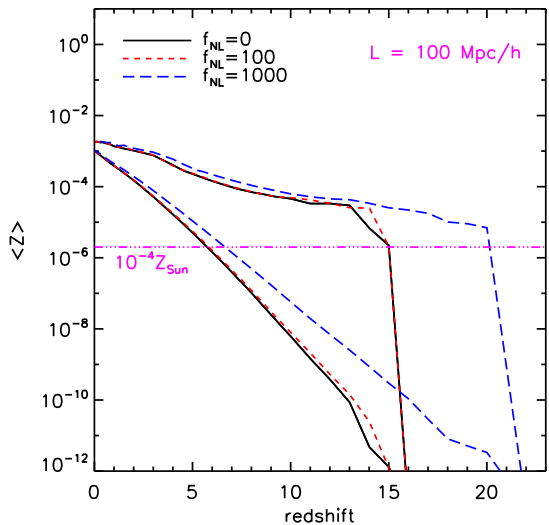}
\caption[Zevolution]{\small
Redshift evolution of the average pollution in the enriched regions (upper lines), and the average pollution in the whole box (lower lines), for \fnl=0 (solid lines), \fnl=100 (short-dashed lines), and \fnl=1000 (long-dashed lines), in the 100~\Mpch{} side boxes.
The horizontal dot-dot-dot-dashed line is drawn at the critical value of $10^{-4}\,Z_\odot$.
}
\label{fig:Zcompare}
\end{figure}
We display, in Fig.~\ref{fig:Zcompare}, the metallicity evolutions for the runs of the \fnl=0, \fnl=100, and \fnl=1000 cases, in the 100~\Mpch{} side box with a top-heavy popIII IMF.
The upper lines correspond to the average enrichment (i.e. the average metallicity of the polluted regions only), and the lower lines to the average metallicity (in the whole simulated volume), as a function of the redshift.
What is particularly evident is the rapidity of early metal pollution \cite[consistently with][]{Maio2010a}.
Indeed, the critical value, $Z_{crit}=10^{-4}\,\rm Z\odot$, is reached in only a few $10^7\,\rm yr$, from the first episodes of star formation and the primordial yields are so high that efficiently pollute the surrounding medium \cite[see more discussion in][]{Maio2011arXiv}.
The following metal evolution shows an increasing trend that leads to an average enrichment of about $Z\sim 10^{-3}-10^{-2}$, i.e. $\sim 0.1\,Z_\odot$.
Due to the patchyness of the metal enrichment process, it is possible to get very strong local pollution, too, with highly supersolar $Z$.
The average metallicity in the whole volume (solid line) increases more gradually, by changing of $\sim 6$ orders of magnitude between $z\sim 16$ and $z\sim 6$.
This behaviour is due to the fact that, as cosmic time evolves, the regions involved in metal pollution are larger and larger, in comparison to the very first isolated events.
However, the average enrichment reaches the critical value, $Z_{crit}$, by redshift $z\sim 6$, when the Universe is roughly one-billion-year old.
Afterwards, during the subsequent $\sim 13\,\rm Gyr$, $Z$ increases of $\sim 3$ orders of magnitude, and the cosmic star formation history is fully dominated by the popII regime (see also Fig.~\ref{fig:SFR}).
Among the different elements, oxygen is the dominant one, since it is heavily produced by PISN and SNII explosions.
While at early times it dominates by about one order of magnitude, at later stages other elements become important, in particular, iron, which is mostly produced by the death of long-living stars.
It catches up carbon and overtakes $\alpha$-elements, like magnesium, sulphur, and silicon.
\\
Briefly speaking, the fact that the average enrichment reaches the critical value already by $z\sim 6$ has broad implications also for reionization, given that metallicity affects the spectral energy distributions (SEDs) of stellar populations \cite[e.g.][]{Schaerer2003}.
In particular, primordial stellar populations have UV fluxes (at wavelengths shorter than $912\,\rm\AA$) which are up to four times larger than the corresponding popII ones, and in principle, could ionize hydrogen in a much more efficient way.
However, given the high level of enrichment in the clustered star forming regions (upper lines in Fig.~\ref{fig:Zcompare}) the popIII star formation rate at $z\sim 6$ is only roughly $\sim 10^{-2}-10^{-4}$ the total one (depending on the modeling).
The relative number fraction of popIII hundred-solar-masses stars with respect to the corresponding popII stars drops further down of a factor of $\sim 10-100$.
As a consequence, it is unlikely that popIII stars will dominate the reionization process simply because they are too rare and, additionally, shine for a too short time (up to few $10^6\,\rm yr$) to provide significant amounts of UV photons.
\\
These conclusions confirm and extend to lower redshift the finding by \cite{Maio2010a}, and show how primordial non-Gaussianities affect the whole picture.
In fact, by comparing the different \fnl{} cases, it emerges that the strongest difference is found for the \fnl=1000 run, in which metal enrichment kicks in at earlier times, tracing back the star formation rate (see Fig.~\ref{fig:SFR}).
Its behaviour is not drammatically different from the other ones, below $z \lesssim 12$, though.
We note that the \fnl=0 and \fnl=100 runs predict extremely similar evolutions, with average metallicities differing by a factor of a few only at the very beginning of the pollution process (at $z\sim 13-15$, where the average enrichment is of $Z<10^{10}$), and rapidly converging later on.
\\
The trends found in all the other runs show very similar behaviour and lead to the same conclusions.


\subsection{Baryon distributions}\label{Sect:distributions}
\begin{figure}
\centering
\includegraphics[width=0.45\textwidth]{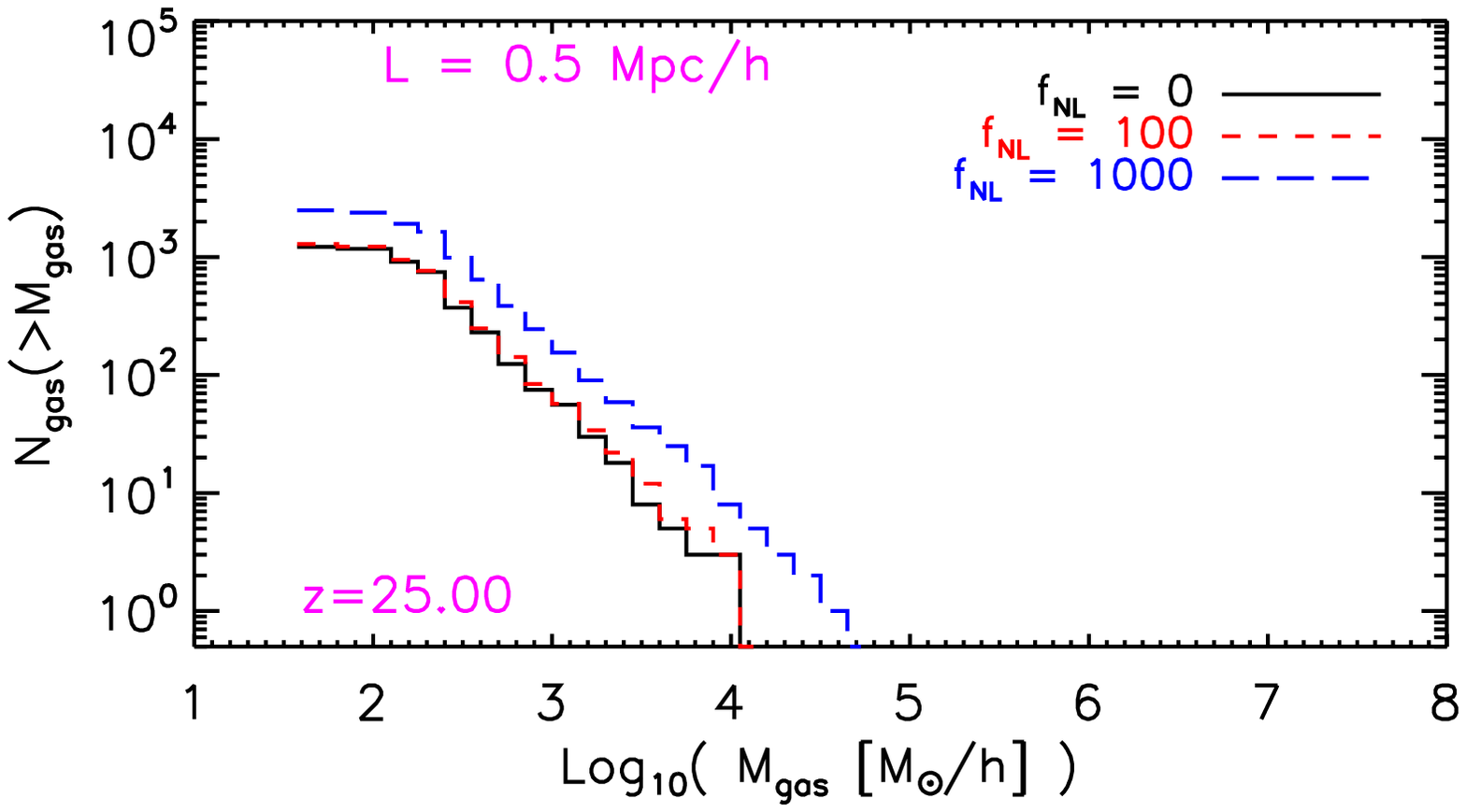}\\
\includegraphics[width=0.45\textwidth]{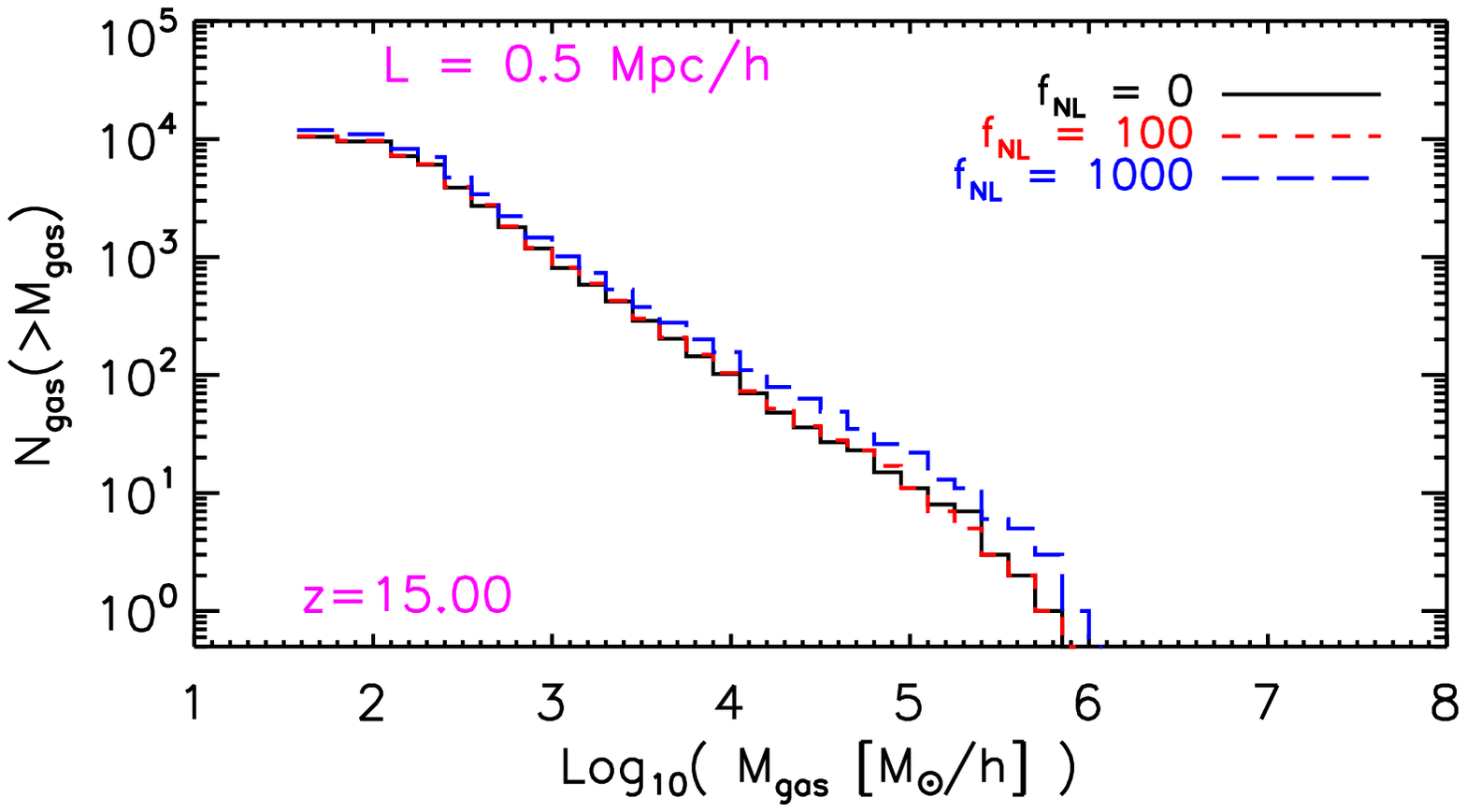}\\
\caption[Ngas]{\small
Cumulative primordial-cloud distributions as a function of gas mass, for $z=25$ (upper panel) and $z=15$ (lower panel), for the runs with \fnl=0 (black solid lines), \fnl=100 (red long-dashed lines), and \fnl=1000 (blue dotted short-dashed lines).
We plot results for the high-resolution simulations with 0.5~\Mpch{} side boxes, labelled as Run05.0, Run05.100, Run05.1000, respectively.
}
\label{fig:Ngas}
\end{figure}
\begin{figure}
\centering
\includegraphics[width=0.45\textwidth]{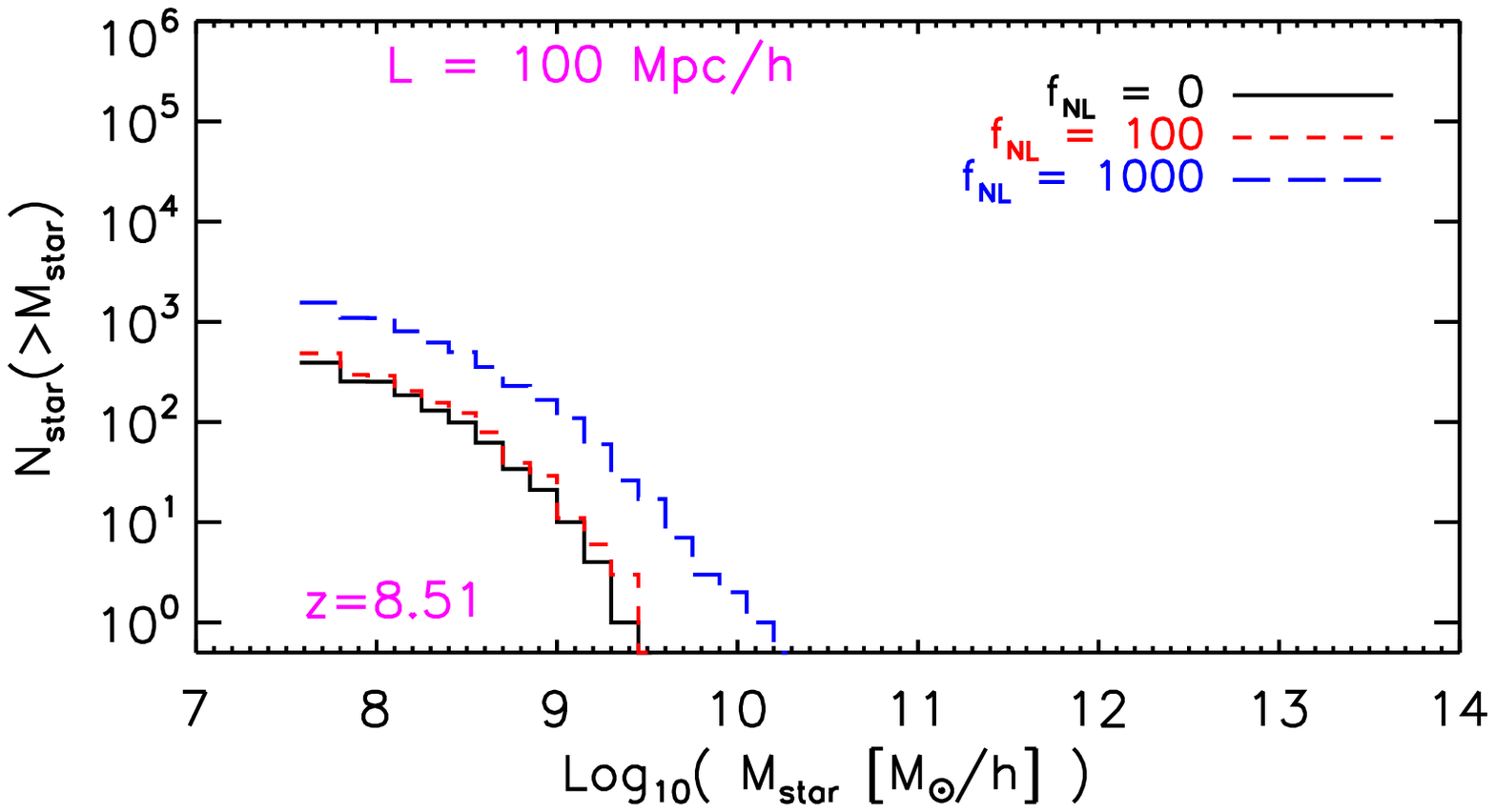}\\ 
\includegraphics[width=0.45\textwidth]{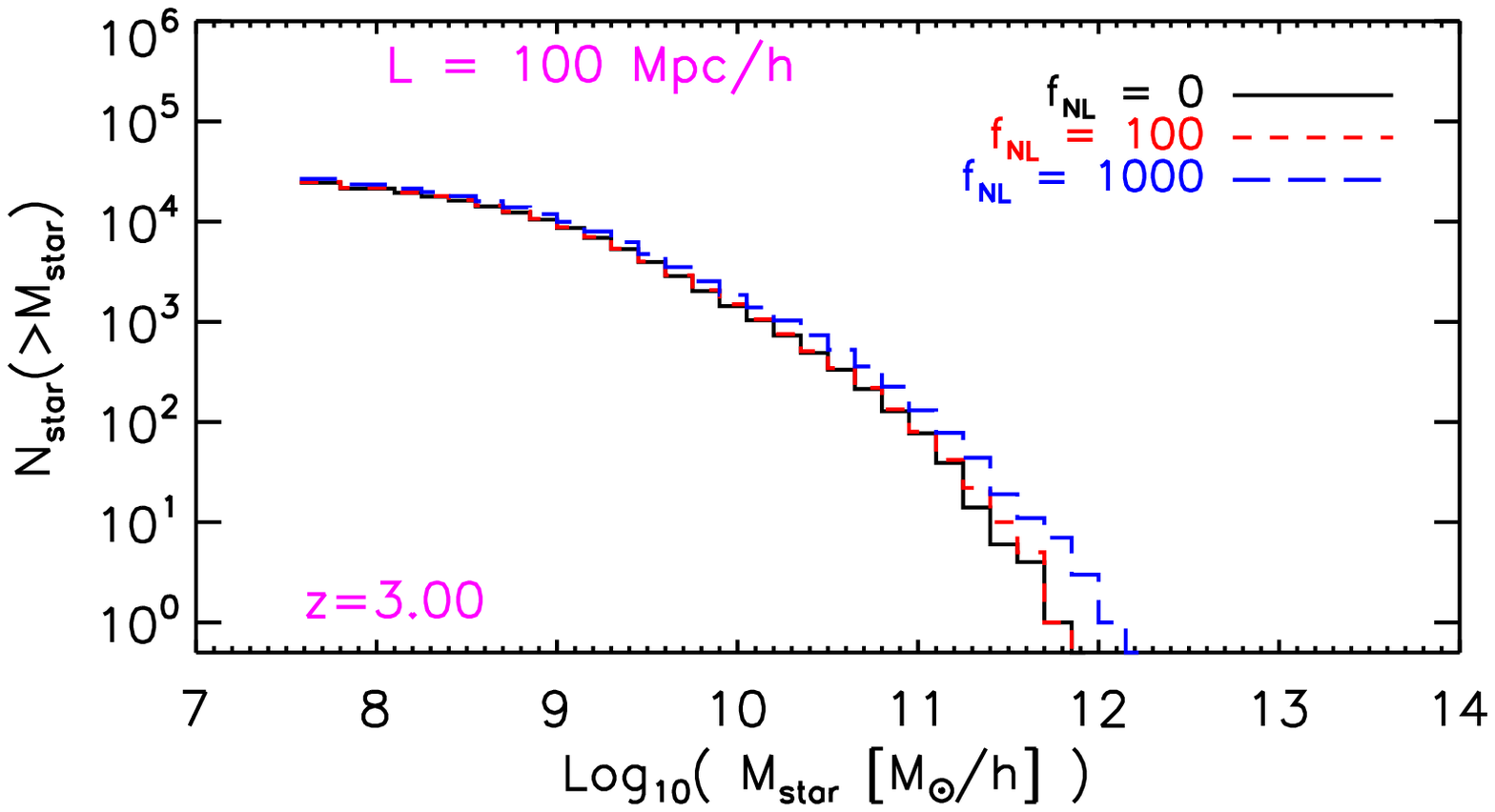}\\ 
\caption[Ngas]{\small
Cumulative cloud distributions as a function of stellar mass, for $z=8.51$ (upper panel) and $z=3.00$ (lower panel), for the runs with \fnl=0 (black solid lines), \fnl=100 (red long-dashed lines), and \fnl=1000 (blue dotted short-dashed lines).
We plot results for the high-resolution simulations with 100~\Mpch{} side boxes, labelled as Run100.0, Run100.100, Run100.1000, respectively.
}
\label{fig:Nstar}
\end{figure}
After discussing the general behaviours of the baryons in the Universe, it is interesting to go through their clumping properties and to investigate how the resulting ``visible'' objects are affected.
So, we identify gas clouds and dark-matter haloes by means of a FoF algorithm, as already described in Sect.~\ref{Sect:simulations}.
\\
To probe the effects of non-Gaussianities on the small primordial gas clouds, in Fig.~\ref{fig:Ngas}, we show the cumulative distributions at early times for the $0.5\,\rm\Mpch$ side boxes.
We focus on the \fnl=0, \fnl=100, and \fnl=1000 cases, at redshift $z=25$ and $z=15$.
At $z\sim 25$, the primordial non-Gaussian perturbations have well visible consequences in the gas clumpiness and lead to more massive and more numerous clouds with respect to the standard Gaussian scenario, for gaseous structures with mass $<10^{7}\msunh$ .
The differences are about a factor of $\sim 2-3$ in the \fnl=1000 case, but only $\sim 10$ per cent for the \fnl=100 case.
This reflects the larger probability at higher \fnl{} of forming objects that are able to trap and condense primordial cosmic gas.
The following growth of structures washes out the differences in a few $\sim 10^8\,\rm Myr$, and by $z\sim 15$ the gas distributions for these small objects are almost undistinguishable.
\\
The following evolution will lead to the formation of protostellar cores, to the birth of the first generation of stars, and then to the establishment of the standard stellar populations (see previous discussion).
To account for the differences which persist down to lower redshift on the stellar evolution properties, we plot in Fig.~\ref{fig:Nstar} the stellar mass distributions in the range $\sim 10^7-10^{13}\msun$, at redshift $z=8.51$ and $z=3$.
We consider the \fnl=0, \fnl=100, and \fnl=1000 cases in the large boxes, because they have higher-mass samples.
We find that also the stellar distributions are affected, mostly at earlier times, with differences that can reach a factor of $\sim 10$ for the \fnl=1000 case and a factor of a few for the \fnl=100 case, in the high-mass tail ($> 10^{9}\,\rm\msunh$).
At the low mass-end ($\sim 10^{7}-10^{8}\,\rm\msunh$), instead, differences are only a factor of $\sim 2$ for \fnl=1000, and $\sim 10$ per cent for \fnl=100.
At low redshift, $z\sim 3$, some differences persist for masses larger than $\sim 10^{11}\,\rm\msunh$, but the most of the objects shows well converging trends.
Below redshift $\sim 3$, the residual differences are further washed out by the subsequent structure formation and the on-going mass assembly process, so at $z\sim 0-1$ there are no distinctions at all in the various stellar components.

\section{Discussion and conclusions}\label{Sect:discussion}

\begin{table}
\centering
\caption[Simulation set-up]{Observational determinations of the \fnl{} parameter, according to different authors and different techniques.}
\begin{tabular}{lcc}
\hline
Reference  & Range for \fnl & Confidence level\\
\hline
\cite{Komatsu2002}  & $[-3500, +2000]$ & $2\sigma$\\
\cite{Komatsu2003}  & $[-58, +134]$    & $2\sigma$\\
\cite{Gaztanaga2003}& $[+4, +18]$      & $1\sigma$\\
\cite{Gaztanaga2003}& $[-4, +5]$       & $1\sigma$\\
\cite{Chen2005}     & $[-260, +40]$    & $1\sigma$\\
\cite{Chen2006}     & $[-30, +74]$     & $1\sigma$\\
\cite{Cabella2006}  & $[-80, +80]$     & $1\sigma$\\
\cite{Cabella2006}  & $[-160, +160]$   & $2\sigma$\\
\cite{Spergel2007}  & $[-54, +114]$    & $2\sigma$\\
\cite{Afshordi2008} & $[+109, +363]$   & $2\sigma$\\
\cite{Slosar2008}   & $[-29, +70]$     & $2\sigma$\\
\cite{Slosar2008}   & $[-65, +93]$     & $3\sigma$\\
\cite{Slosar2008}   & $[-31, +70]$     & $2\sigma$\\
\cite{Slosar2008}   & $[-96, +96]$     & $3\sigma$\\
\cite{Hikage2008}   & $[-70, +91]$     & $3\sigma$\\
\cite{Yadav2008}    & $[+27, +147]$    & $2\sigma$\\
\cite{Komatsu2009}  & $[-9,  +111]$    & $2\sigma$\\
\cite{Smith2009}    & $[-4,  +80]$     & $2\sigma$\\
\cite{Curto2009}    & $[-8,  +111]$    & $2\sigma$\\
\cite{Curto2009b}   & $[-18, +80] $    & $2\sigma$\\
\cite{Vielva2009}   & $[-94, +154] $   & $2\sigma$\\
\cite{Rudjord2009}  & $[+44, +124]$    & $2\sigma$\\
\cite{Rudjord2010}  & $[+42, +104]$    & $2\sigma$\\
\cite{Hou2010}      & $[+34, +120]$    & $1\sigma$\\
\cite{Hou2010}      & $[+12, +82]$     & $1\sigma$\\
\cite{Cabella2010}  & $[-9, +85]$      & $1\sigma$\\
\cite{Komatsu2010}  & $[+11, +53]$     & $1\sigma$\\
\cite{Komatsu2010}  & $[-10, +74]$     & $2\sigma$\\
\hline
\label{tab:nG}
\end{tabular}
\begin{flushleft}
\vspace{-0.5cm}
{\small
}
\end{flushleft}
\end{table}
Cosmological structure formation is strongly affected by the initial perturbations which it originates from.
As a consequence of the central-limit theorem, it is common to assume a Gaussian distribution, but detailed studies of the CMB leave some room for deviations from the expected behaviour, and for an \fnl{} parameter different from zero.
Since non-Gaussianities alter the ``initial conditions'' of the problem of structure formation, and, thus, influence the temporal evolution of the forming structures, we have addressed their impacts on the visible, baryonic matter.
\\
Given the lack of extensive works on this subject, we have performed the first high-resolution cosmological N-body, hydrodynamical, chemistry simulation of cosmic structure formation, in the frame of the $\Lambda$CDM model and in presence of primordial non-Gaussianities.
We have studied the principal properties of gas and stars in different scenarios, by following: cosmic evolution of e$^-$, H, H$^+$, H$^-$, He, He$^+$, He$^{++}$, H$_2$, H$_2^+$, D, D$^+$, HD, HeH$^+$ \cite[][]{Yoshida2003,Maio2006,Maio2007,Maio2009,Maio2010a}; radiative gas cooling at high \cite[][]{SD1993} and low temperatures \cite[both from molecules and fine-structure metal transitions, according to ][]{Maio2007}; multiphase star formation \cite[][]{Springel2003}; UV background radiation \cite[][]{HaardtMadau1996}; wind feedback \cite[][]{Springel2003,Aguirre_et_al_2001}; metal pollution from popIII and/or popII stellar generations, ruled by a critical metallicity threshold of $Z_{crit}=10^{-4}\,\zsun$ \cite[][]{Tornatore2007,Maio2010a} and leading gravitational enrichment into the surrounding medium \cite[][]{Maio2011arXiv}.
\\
Observational evidences and continuous investigations over the last decade (see details in Tab.~\ref{tab:nG}) have progressively constrained \fnl{} to positive values, thus, we have considered \fnl=0, 10, 50, 100, and 1000.
From Tab.~\ref{tab:nG}, the most likely values\footnote{
In Tab.~\ref{tab:nG}, the CMB notation is assumed, thus, to compare with the values adopted in the simulations (\fnl=0, 10, 50, 100, 1000), one has to multiply the tabulated data by $\sim 1.28$.
}
 are expected to be between \fnl$\sim 0-100$, and the \fnl=1000 case has to be simply considered as an extreme upper limit.
\\
The goal of this work has been to theoretically explore the effects of different \fnl{} values on the structure formation properties and on the baryon history.
Thanks to the high resolution reached (up to $\sim 40\,\msunh$), we are able to address the cosmic evolution from the very early times -- when the first star formation episodes take place -- till to the present.
\\
Our main results (Sect.~\ref{Sect:results}) have shown that structure formation processes (Sect.\ref{Sect:maps}), star formation history (Sect.~\ref{Sect:SFR}), chemical evolution (Sect.~\ref{Sect:chemistry}), and visible-matter distributions (Sect.~\ref{Sect:distributions}) are all affected by non-Gaussianities.
However, while the effects are quite evident for the extreme \fnl=1000 case, they are much milder for smaller, more ralistic values of \fnl.
\\
Indeed, gas temperature and composition reveal some effects of primordial non-Gaussianities only at very high redshift, with changes of the order of $\sim 10$ per cent, at $z > 10$, for \fnl=$0-100$ and of a factor of $\sim 2$ for \fnl=1000 (Fig.~\ref{fig:Maps}).
\\
Primordial molecular fractions  (Fig.~\ref{fig:Maps2} and Fig.~ \ref{fig:Maps3}) are particularly senstive to the thermodynamical properties of the medium and their behaviour increases mostly in the high-density regions, where the effects on clustering due to primordial non-Gaussianities are more relevant.
\\
Star formation history (Fig.~\ref{fig:SFR}) has a converging behaviour despite of non-Gaussianities. The main discrepancies among the models are found for the onset, which takes place at earlier epochs for larger \fnl.
The time-lags are around some $10^7\,\rm yr$, at $z\sim 15$, and they are reflected in the transition from the popIII regime, described by a top-heavy IMF, to the popII regime, described by a Salpeter IMF (Fig.~\ref{fig:SFR}).
Despite such discrepancies, the popIII contributions stay around $\sim 10^{-3}-10^{-2}$ at $z\sim 10$ for all the models, with only $\sim 1-10$ per cent scatter.
\\
Given the uncertainties on the primordial popIII IMF we also run corresponding simulations with Salpeter-like popIII IMF and found that (Fig.~\ref{fig:SFR_IMFrange}) the resulting deviations of the popIII star formation rates could be larger than the non-Gaussian ones.
\\
Additionally, we compared with the second-order corrections in linear perturbation theory \cite[][]{TseliakhovichHirata2010} and their impact on structure formation evolution \cite[][]{Maio2011}, showing that (Fig.~\ref{fig:SFR_vb}) also the resulting primordial gas bulk flows could have comparable or larger repercursions on the visible matter.
Therefore, the behaviour of baryons seems tightly bound and twisted with several physical mechanisms, whose effects are degenerate with non-Gaussianities.
\\
As a consequence of the trends of the star formation rates, metal enrichment results (Fig.~\ref{fig:Zcompare}) are affected by the initial \fnl, mostly at early times, with pollution histories converging at low redshift ($z\lesssim 12$) and overcoming, in average, the critical metallicity at $z\sim 6$, when the Universe has an age of roughly 1~Gyr.
This is quite independent from the primordial non-Gaussianities, popIII IMF, or additional second-order effects, but at high redshift metal spreading is highly sensitive to all these phenomena and can show degeneracies with them.
\\
Gaseous and stellar mass statistics (Fig.~\ref{fig:Ngas}, and Fig.~\ref{fig:Nstar}) show variations at the high-mass tail, and rare, more massive objects are more common in higher-\fnl{} cosmologies.
The distributions of the very first structures show an increment of a factor of a few in the \fnl=1000 case, but only some $\sim 10$ per cent for \fnl$=0-100$, mainly at the high-mass end.
However, the subsequent evolution at later times washes them out almost completely.
\\
Throughout this work, we have used the \fnl{} parameter only to quantify deviations from the Gaussian shape.
We have not investigated models with a negative \fnl{} as they have been progressively dismissed by observations.
However, for sake of completness, we note that, theoretically, one would expect roughly specular behaviours with respect to the \fnl=0 case when universes with negative \fnl{} are considered -- a result already well established from simple dark-matter-only simulations (as mentioned in Sect.~\ref{Sect:introduction}).
Indeed, in such scenarios, the probability distribution function of the initial density field would be skewned towards the underdense tail, causing a paucity of the more massive objects and, in general, a delayed evolution of the cosmic structures.
We therefore expect the variations from the \fnl=0 case originated from a positive \fnl{} parameter to be accordingly reversed when the corresponding negative \fnl{} parameter is adopted.
\\
Besides the \fnl{} formulation, higher-order parameters can be introduced, but their uncertainties are huge with, e.g., $-7.4\times 10^5<$\gnl$<8.2\times 10^5$ \cite[][]{Smidt2010}.
Thus, a study of their impacts on the cosmic structures would be quite vague and elusive.
Furthermore, the effects of the second-order \fnl{} parameter are quite small, so it is reasonable to think that inclusion of higher-order parameters will not dramaticaly change our conclusions.
Even if future initial conditions will include additional higher-order corrections, being definitely more complete and detailed, probably they would not alter the global scenario we have drawn.
For what we said in the previous sections, other physical phenomena, like 
cosmic rays\footnote{
See e.g. \cite{Ginzburg1964,Mannheim1994,Biermann2001,Sigl2004,Stanev2004,Vasiliev2006,Pfrommer2007,Jasche2007}.
},
radiative transfer and reionization\footnote{
See e.g. \cite{Abel1999,Gnedin2001,Bene2001,Bene2003,Maselli2003,Whitehouse2005,Iliev2006,Mellema2006,Trac2007,Aubert2008,Johnson2008,Pawlik2008,Finlator2009,Margarita2009,Margarita2010,Cantalupo2011,Wise2010arXiv}.
},
magnetic fields\footnote{
See e.g. \cite{Orszag1979,Phillips1985,Brio1988,Dolag1999,Balsara1999,Londrillo2000,Toth2000,Borve2001,Brueggen2005,Price2005,Li2008,Stone2008,Ryu2008,Dolag2009,Brandenburg2010,Burzle2010arXiv,Gaburov2010arXiv,Kotarba2010}.
},
AGN feedback\footnote{
See e.g. \cite{Begelman1978,Blandford1982,Rees1984,Ciotti1997,King2003,Sazonov2005,Churazov2005,Granato2006,Ciotti2007,Khochfar2008,DiMatteo2008,Sijacki2009,Milo2009,McKernan2010,Volonteri2010arXiv,Teyssier2010arXiv,McNamara2011}.
},
etc., are not supposed to experience strong modifications, because of primordial non-Gaussianities.
\\
We stress that even small variations in the determinations of the cosmological parameters, above all errors of a few per cent in the estimates of $\sigma_8$, could add further degeneracies to disentangle the purely non-Gaussian results from the primordial IMF, gas bulk flows, etc.
\\
We summarize our conclusions by stating that for reasonable values of \fnl$=0-100$ in the primordial non-Gaussianities (in agreement with observations):
\begin{itemize}
\item
  early molecular fractions in the cold, dense gas phase are predicted to be altered of $\sim 10$ per cent;
\item
  small temperature fluctuations of $\,\lesssim 10$ per cent are induced during the first Gyr of the cosmic evolution;
\item
  the onset of the first star formation events, at $z\gtrsim 15$, is influenced of some $10^7\,\rm yr$;
\item
  the epoch of popIII/popII transition is affected by some $10^7\,\rm yr$;
\item
  variations of $\lesssim 10$ per cent are found in the mass distributions during the formation of first structures;
\item
  only mild variations in the chemical history of the Universe are expected;
\item
  extremly high values for \fnl{} would imply corresponding variations of a factor $\sim 2-3$ in the thermodynamical and chemical properties of the gas, earlier star formation histories of $\sim 0.5-1\times 10^8\,\rm yr$, and changes up to a factor of $\sim 10$ in the high-mass end of the gas cloud and stellar mass distributions at early times;
\item
  non-Gaussian effects could have degeneracies with other physical processes, like primordial gas bulk flows, with unknown star formation quantities, like popIII IMF, or with the determination of cosmological parameters, mostly $\sigma_8$ at a few per cent level.
\end{itemize}


\section*{acknowledgments}
We acknowledge useful discussions with K.~Dolag, M.~Grossi, and L.~Moscardini.
UM also acknowledges B.~Ciardi, S.~Khochfar, and G.~Rossmanith.
We also thank the referee M.~Stiavelli for his useful and prompt comments on the manuscript.
The simulations were performed on the IBM Power~6 machines of the Garching Computing Center (Rechenzentrum Garching) of the Max Planck Society (Max-Planck-Gesellschaft).


\bibliographystyle{mn2e}
\bibliography{bibl.bib}

\label{lastpage}
\end{document}